\documentclass{aa}
\usepackage[varg]{txfonts}
\usepackage{graphicx}	
\usepackage{amsmath}	
\usepackage{subfigure}
\usepackage{epstopdf}
\usepackage[colorlinks, linkcolor=blue, urlcolor=blue, citecolor=blue]{hyperref}
\def\beq{\begin{equation}}
\def\enq{\end{equation}}
\def\bea{\begin{eqnarray}}
\def\ena{\end{eqnarray}}

\begin{document}

\title{New Insights into the Criteria of Fast Radio Burst in the Light of FRB 20121102A}

\author{Di Xiao\inst{\ref{inst1},\ref{inst2}}
\and Zi-Gao Dai \inst{\ref{inst3},\ref{inst1}}}

\institute{School of Astronomy and Space Science, Nanjing University, Nanjing 210023, China \label{inst1} \\
\email{dxiao@nju.edu.cn}
\and
Key Laboratory of Modern Astronomy and Astrophysics (Nanjing University), Ministry of Education, China \label{inst2}
\and
Department of Astronomy, University of Science and Technology of China, Hefei 230026, China \label{inst3}}

\date{Received XXX / Accepted XXX}
\abstract{The total event number of fast radio bursts (FRBs) is accumulating rapidly with the improvement of existing radio telescopes and the completion of new facilities. Especially, the Five-hundred-meter Aperture Spherical radio Telescope (FAST) Collaboration has just reported more than one thousand bursts in a short observing period of 47 days \citep{LiD2021}. The interesting bimodal distribution in their work motivates us to revisit the definition of FRBs. In this work, we ascribe the bimodal distribution to two physical kinds of radio bursts, which may have different radiation mechanisms. We propose to use brightness temperature to separate two subtypes. For FRB 20121102A, the critical brightness temperature is $T_{\rm B,cri}\simeq10^{33}\,\rm K$. Bursts with $T_{\rm B}\geq T_{\rm B,cri}$ are denoted as ``classical'' FRBs, and further we find a tight pulse width-fluence relation ($T\propto\mathcal{F_\nu}^{0.306}$) for them. On the contrary, the other bursts are considered as ``atypical'' bursts that may originate from a different physical process. We suggest that for each FRB event, a similar dividing line should exist but $T_{\rm B,cri}$ is not necessarily the same. Its exact value depends on FRB radiation mechanism and properties of the source.}
\titlerunning{Classification of FRBs by brightness temperature}
\authorrunning{D. Xiao and Z. G. Dai}
\maketitle
\section{Introduction}
\label{sec1}
Fast radio bursts (FRBs) are new bright millisecond radio pulses that the first discovery was just in 2007 \citep{Lorimer2007}. In early years, doubt exists whether FRBs are astrophysical, until a new sample has been identified in 2013 \citep{Thornton2013}. Ever since then this mysterious phenomenon starts to attract intense attention from the community with breakthroughs coming one after another \citep[for reviews, see][]{Katz2018, Popov2018, Petroff2019,Cordes2019,Zhang2020c,XiaoD2021,Petroff2021}. One striking achievement is the discovery of the first repeating event FRB 20121102A in 2016 \citep{Spitler2016,Scholz2016} and later the identification of its host galaxy \citep{Chatterjee2017,Marcote2017,Tendulkar2017}. Up to now, there are more than twenty repeaters and six hundred apparently non-repeating FRBs in the catalog \citep{Petroff2016,CHIME2021cat}. However, it is still under hot debate whether genuinely non-repeating FRBs exist \citep{Caleb2018b,Palaniswamy2018,Caleb2019,XiaoD2021}. A few works have discussed the possibility to judge this question using the number fraction of repeaters \citep{Caleb2019, Lu2020a, Ai2021,Gardenier2021}. With the accumulation of observing time, this fraction will increase to 100 \% if all FRBs repeat. Otherwise, it will peak at a value less than 100 \% at a certain time \citep{Ai2021}.

Except for the repeating behavior, currently there is no good criterion for the classification of FRBs yet. Actually, the definition for FRBs is not very strict now. Generally, an FRB is characterized by duration ($T\sim$ millisecond) and extremely high brightness temperature (typically $T_{\rm B}>\,10^{30}\,\rm K$). Sometimes a large dispersion measure is needed to distinguish FRBs from rotating radio transients (RRATs). Among all the observational properties, brightness temperature seems the most promising criterion for the classification because it relates to the radiation mechanism directly. Different coherent radio emission mechanisms should have their own ``extremes'', resulting in various transient phenomena such as pulsar radio emission, RRATs, giant pulses, nanoshots and so on. They cluster in different regions of the spectral luminosity-duration phase space for radio transients, and the most prominent difference between them is $T_{\rm B}$ \citep[e.g., Fig. 3 of][]{Nimmo2021}. However, the critical value $T_{\rm B,cri}$ to define an FRB is not well known. Very recently, the large sample of FRB 20121102A bursts released by the Five-hundred-meter Aperture Spherical radio Telescope (FAST) Collaboration makes it possible to explore this issue \citep{LiD2021}.

FRB 20121102A has been well studied before and observed by different radio telescopes \citep{Spitler2016,Gourdji2019,Hessels2019,Rajwade2020,Cruces2021}. Its burst rate is very high, thus could be a nice event for studying FRB classification. Previous works showed that the energy distribution can usually be approximated by a power-law form, but the index varies while using different samples from different telescopes \citep{WangFY2017,Law2017,Gourdji2019,WangFY2019,ZhangGQ2021}. Similar distributions have been found for solar type III radio bursts \citep{WangFY2021}, as well as magnetar bursts \citep{Gogus1999,Prieskorn2012,WangFY2017,ChengYJ2020}, which are considered as close relatives to FRBs. However, as the largest single-dish radio telescope now, FAST has a very high sensitivity and the energy threshold for detection is lower than ever. The new sample of FRB 20121102A consists of 1652 bursts and the rate is as high as $\sim122\, \rm hr^{-1}$. Intriguingly, a bimodal burst energy distribution has been found \citep{LiD2021}, indicating that there are two subtypes of FRBs. This motivates us to consider the criterion for FRBs, and we expect to find some empirical relation in the classified subtype sample.

This letter is organized as follows. We introduce the method of classification by brightness temperature and apply it to the FAST sample in Section \ref{sec2}. An evident two-parameter empirical relation is found in Section \ref{sec3} and we discuss the reasonability of choosing the value of $T_{\rm B, cri}\simeq10^{33}\,\rm K$ for FRB 20121102A. We finish with discussion and conclusions in Section \ref{sec4}.

\section{FRB classification by brightness temperature}
\label{sec2}

\begin{figure}
	\begin{center}
		\includegraphics[width=0.42\textwidth]{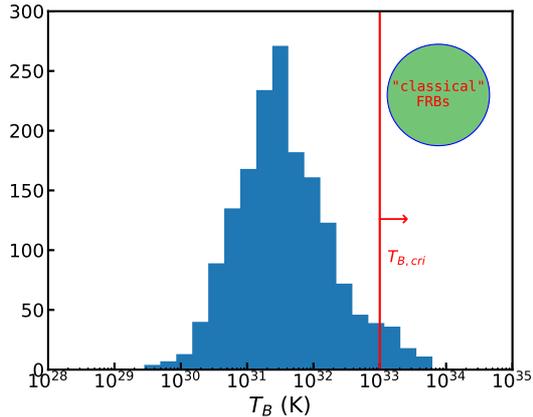}
		\caption{The distribution of $T_{\rm B}$ for FAST 1652 bursts. Red vertical line represents the critical brightness temperature of defining ``classical'' FRBs.}
		\label{fig1}
	\end{center}
\end{figure}

The brightness temperature of an FRB is determined by equaling the observed intensity with blackbody luminosity, which gives
\bea
T_{\rm B}&=&F_\nu d_{\rm A}^2/2\pi k (\nu T)^2\nonumber\\
&=&1.1\times10^{35}{\,\rm K}\,\left(\frac{F_\nu}{\rm Jy}\right)\left(\frac{\nu}{\rm GHz}\right)^{-2}\left(\frac{T}{\rm ms}\right)^{-2}\left(\frac{d_{\rm A}}{\rm Gpc}\right)^2,\nonumber\\
\label{eq:T_B}
\ena
where $F_\nu$ is flux density, $\nu$ is the emission frequency and $T$ is the pulse width. Note that it should be angular diameter distance $d_{\rm A}$ here instead of luminosity distance $d_{\rm L}$ \citep{Zhang2020c,XiaoD2021}, therefore the brightness temperatures of FRBs have been overestimated in previous works. Adopting the redshift of $z=0.19273$ \citep{Tendulkar2017}, we get $d_{\rm A}=0.682\,\rm Gpc$ for FRB 20121102A using the cosmological parameters $H_0=67.7\,\rm km\,s^{-1}\,Mpc^{-1}$, $\Omega_m=0.31$, $\Omega_\Lambda=0.69$ \citep{Planck2016}. With the burst properties given in Supplementary Table 1 of \citet{LiD2021}, we can easily obtain $T_{\rm B}$ for each of the 1652 bursts and its distribution is shown in Figure \ref{fig1}. It is obvious that this distribution centers around $\sim \,\rm a\,\, few\times10^{31} K$, lower than the typical $T_{\rm B}$ of other FRB events. This can be attributed to the ability of detecting low-energy bursts by FAST.

\begin{figure}
	\begin{center}
		\includegraphics[width=0.42\textwidth]{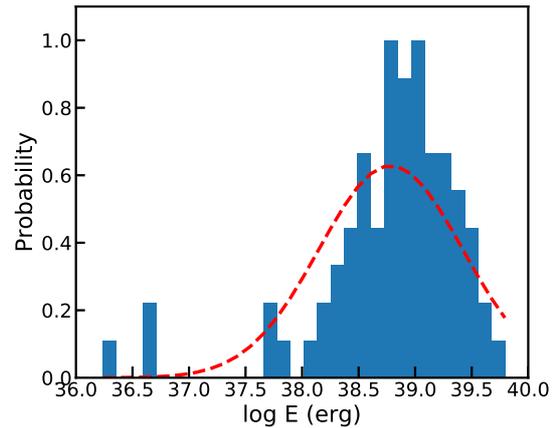}
		\caption{The energy distribution of ``classical'' subtype of the FAST bursts. Red dashed line shows the probability density function of the Gaussian distribution.}
		\label{fig2}
	\end{center}
\end{figure}

However, the main point we argue is that there is a critical line $T_{\rm B,cri}$, so the majority of these low energy bursts may be ``atypical''. Bursts with $T_{\rm B}\geq T_{\rm B,cri}$ are considered as ``classical'' FRBs. Since this critical value is unknown, tentatively we choose $T_{\rm B,cri}=10^{33} \,\rm K$ indicated by the plot of spectral luminosity-duration phase space \citep{Keane2018,Nimmo2021,Petroff2021}. The influence of choosing different $T_{\rm B,cri}$ will be discussed later in Section \ref{sec3}. After drawing this dividing line, we find 76 out of 1652 bursts are ``classical''. Further the energy distribution for this subtype is shown in Figure \ref{fig2}. Red dashed line represents the probability density function of Gaussian distribution, which has the form
$f(x)=\frac{1}{\sigma\sqrt{2\pi}}\exp\left[-\frac{1}{2}\left(\frac{x-\mu}{\sigma}\right)^2\right],$
where $\mu$ is the mean value and $\sigma$ is standard deviation. We can see the energy distribution is consistent with a single Gaussian profile now. Our result is different from an another latest analysis of 1652 FAST bursts that suggested a power-law energy distribution for high-energy bursts \citep{ZhangGQ2021}. This may arise from the different method of classification and they used burst energy instead.  Till now it might be still not convincing enough to classify FRBs by $T_{\rm B}$, however, further support can be obtained if we find some similarity within these subtype bursts.

\section{A two-parameter relation for ``classical'' bursts of FRB 20121102A}
\label{sec3}
\begin{figure}
	\centering
	\subfigure{
		\begin{minipage}[b]{0.45\textwidth}
			\includegraphics[width=1\textwidth]{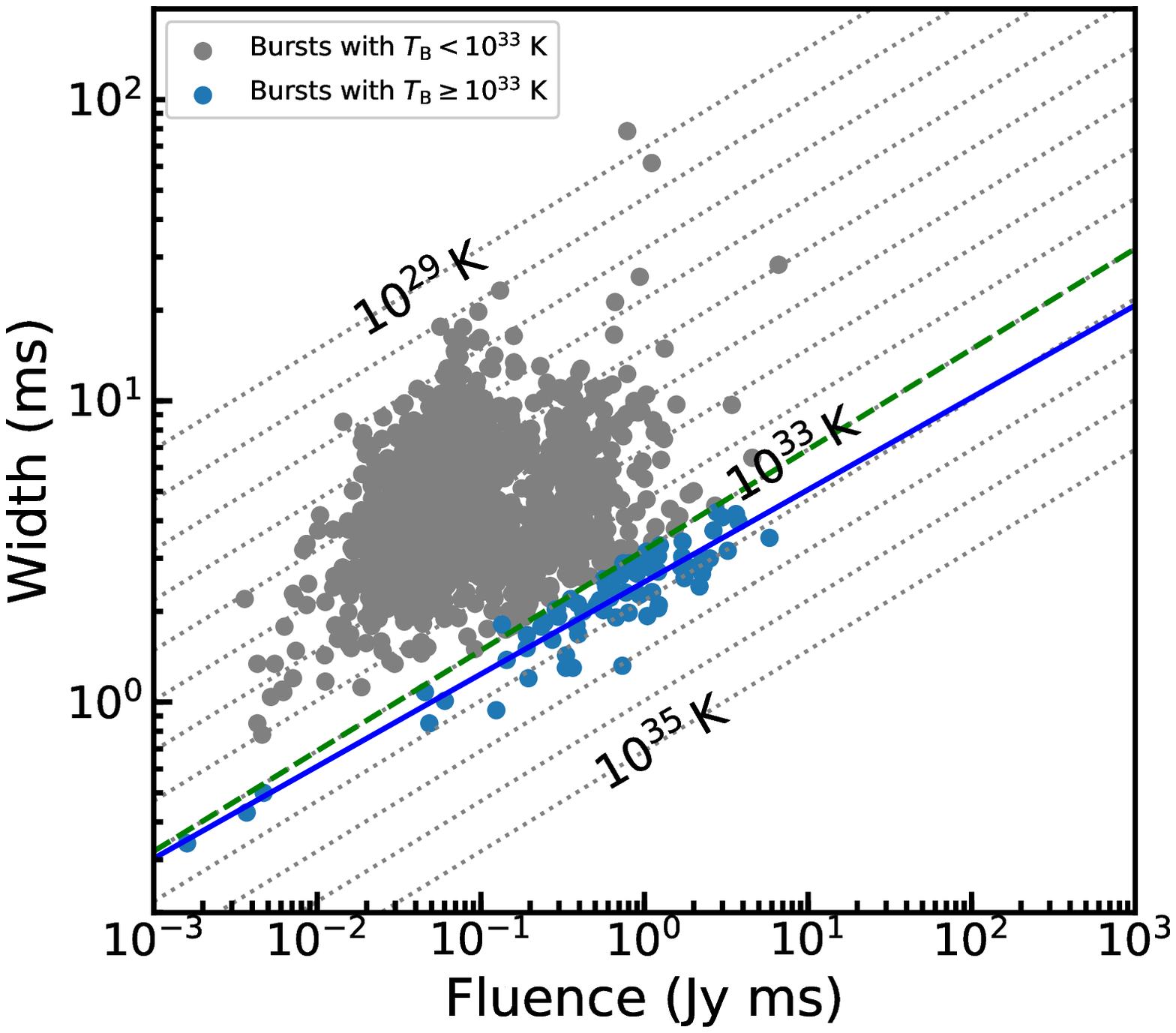}
			\label{fig3:subfig:a}
		\end{minipage}
	}
	\subfigure{
		\begin{minipage}[b]{0.45\textwidth}
			\includegraphics[width=1\textwidth]{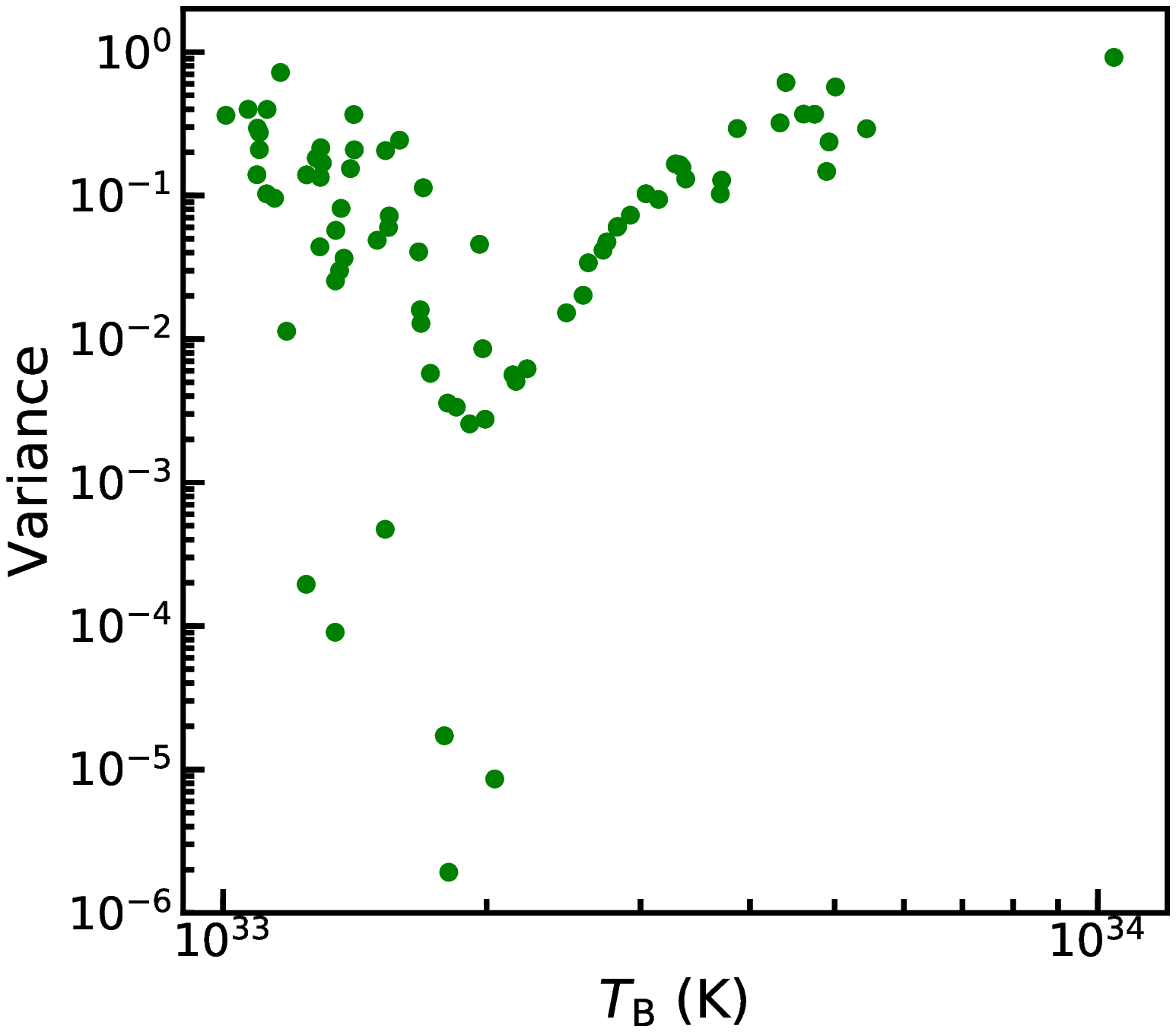}
			\label{fig3:subfig:b}
		\end{minipage}
	}
	\caption{Upper Panel: The scatter plot of pulse width versus fluence for the full FAST sample, in which 76 ``classical'' bursts are marked with colored dots. The blue solid line represents the best-fitting of $T\propto\mathcal{F_\nu}^{0.306}$. The dotted lines represent for different fixed $T_{\rm B}$ values and the green dashed line is for the critical $10^{33}\,\rm K$. Lower Panel: The variance of 76 bursts with respect to the fitting line versus their brightness temperatures. No clear dependence is found.}
	\label{fig3}
\end{figure}

As we have pointed out, if the subtype of ``classical'' bursts is true, they should all originate from one particular physical mechanism and have some properties in common. We examine whether a correlation exists between burst width $T$ and fluence $\mathcal{F_\nu}$. Intriguingly, while the full FAST sample looks rather scattered on the $T-\mathcal{F_\nu}$ plane, we find a tight relation for these 76 ``classical'' bursts, which is shown in the upper panel of Figure \ref{fig3}. The best-fitting blue line is
\bea
\log \left(\frac{T}{\rm ms}\right)=0.306\log\left(\frac{\mathcal{F_\nu}}{\rm Jy\,ms}\right)+0.399.
\ena
We have calculated the correlation coefficient commonly used in linear regression analysis by $ r=\sqrt{{\sum_k(\hat{y}_k-\bar{y})^2}/{\sum_k(y_k-\bar{y})^2}}, $
where $y_k, \,\hat{y}_k,\,\bar{y}$ are the observed value, regressed value and mean of observed value respectively. The correlation coefficient is $r=0.936$, very close to unity. Furthermore, an F-test is implemented, where the method of calculating $F_{\rm value}$ can be found in \citet{TuZL2018}. We obtain that $F_{\rm value}=527.63$ for these 76 bursts, much larger than the critical $F_{\alpha,1,76-2}=3.97$ where $\alpha=5\%$ is adopted. Therefore the null hypothesis is rejected and the relation $T\propto\mathcal{F_\nu}^{0.306}$ is well established. Since the faint bursts generally have low brightness temperatures (see subsequent Figure \ref{fig6}) and seem to show complex time-frequency structures \citep{LiD2021}, we plot the dependence of variance on $T_{\rm B}$ in the lower panel of Figure \ref{fig3} but find no clear correlation. Therefore, the morphological changes of the bursts do not have a prominent effect on the $T-\mathcal{F_\nu}$ relation.

\begin{figure}
	\centering
	\subfigure{
		\begin{minipage}[b]{0.45\textwidth}
			\includegraphics[width=1\textwidth]{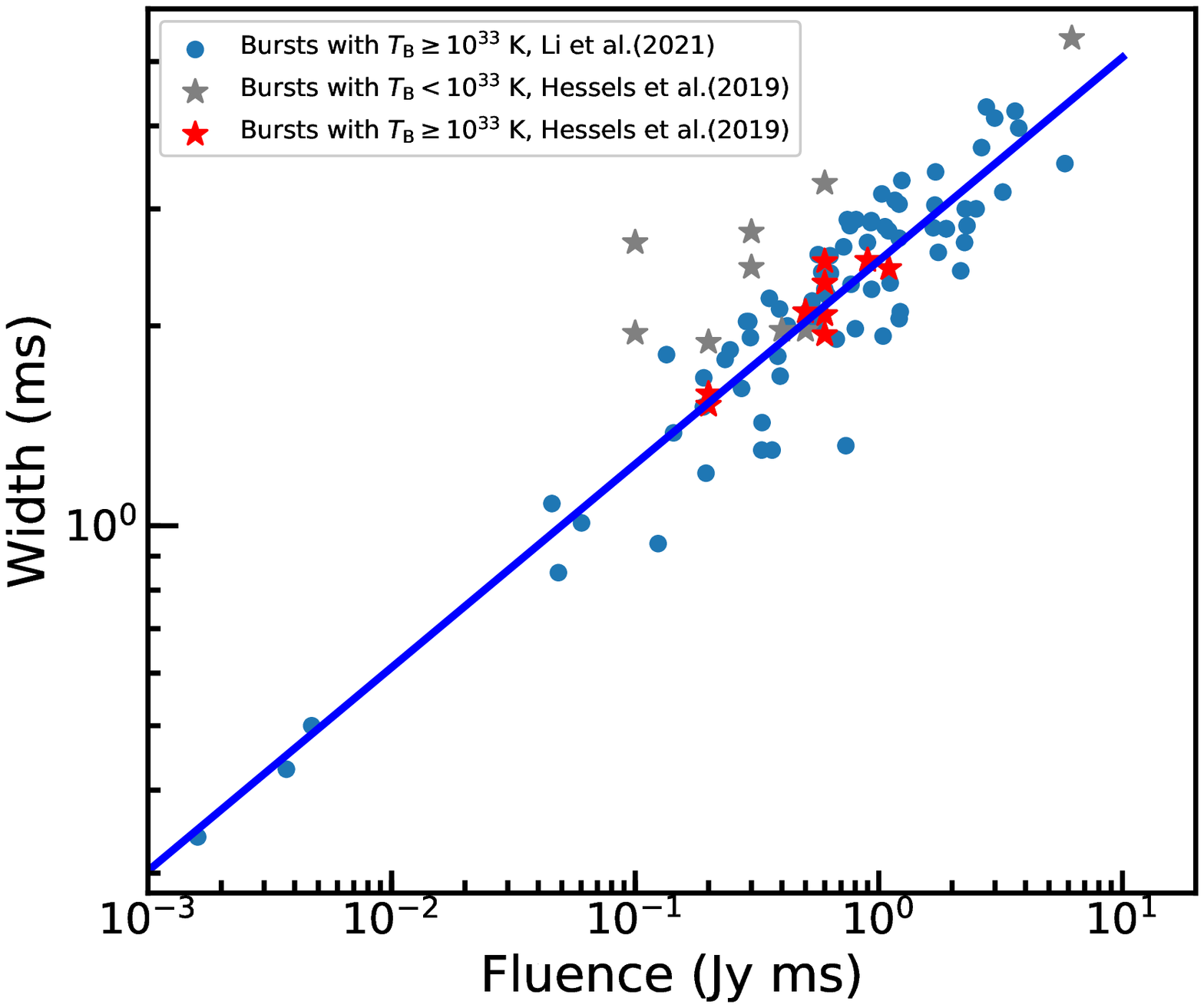}
			\label{fig4:subfig:a}
		\end{minipage}
	}
	\subfigure{
		\begin{minipage}[b]{0.45\textwidth}
			\includegraphics[width=1\textwidth]{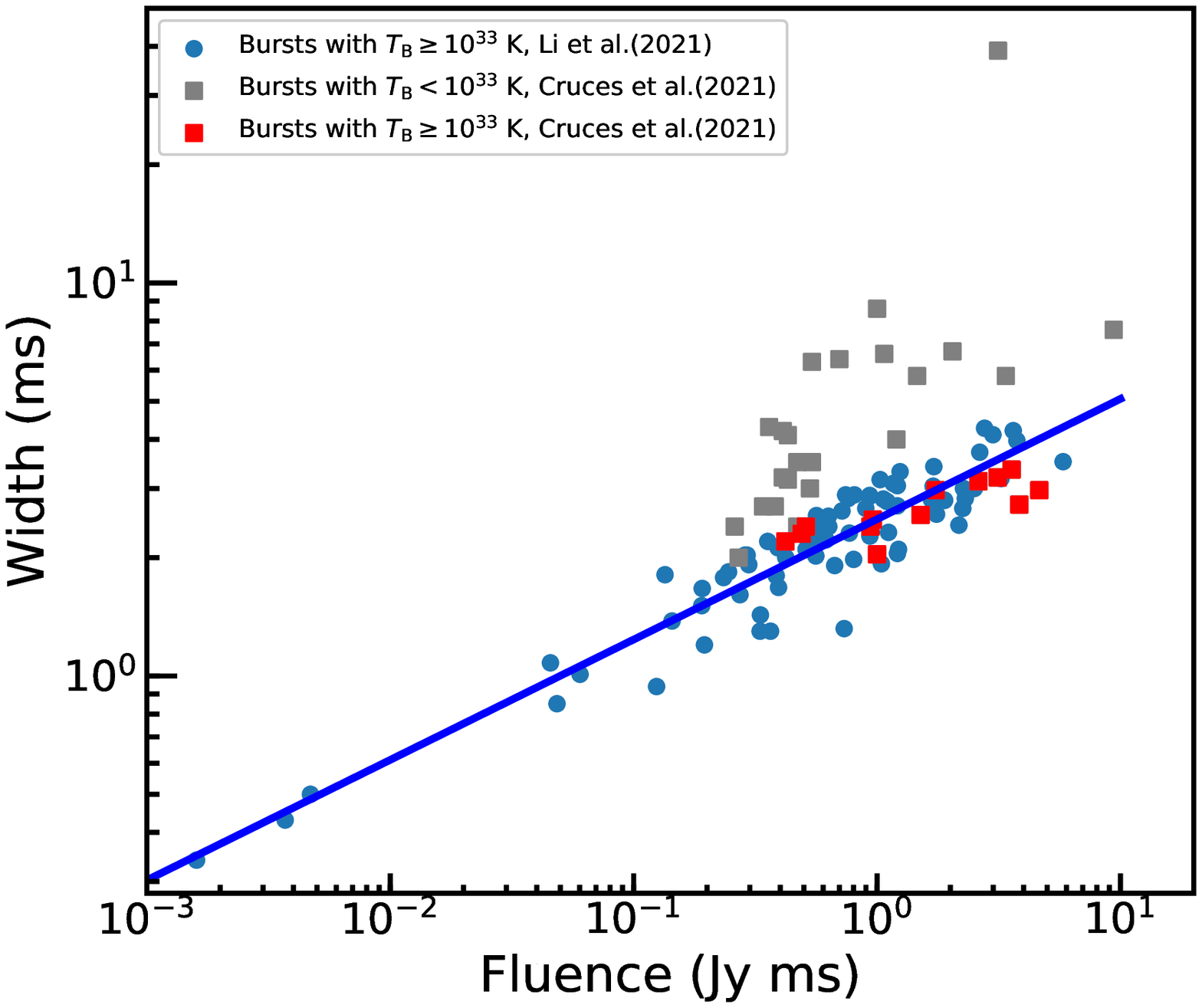}
			\label{fig4:subfig:b}
		\end{minipage}
	}
	\caption{The scatter plot of pulse width versus fluence for two other samples of FRB 20121102A. Blue dots and line are the same as in Figure \ref{fig3}. Upper panel: 19 bursts from \citet{Hessels2019}, marked by stars. Lower Panel: 36 bursts from \citet{Cruces2021}, marked by squares. In both panels the red ones are ``classical'' bursts while the grey ones are ``atypical''. Clearly the red ones are consistent with the $T\propto\mathcal{F_\nu}^{0.306}$ relation.}
	\label{fig4}
\end{figure}

In order to verify this relation, randomly we have analyzed two other samples of FRB 20121102A, i.e., 19 bursts detected by Arecibo and GBT \citep{Hessels2019} and 36 bursts by Effelsberg \citep{Cruces2021}. Different telescopes has different central frequencies and we repeat the above procedure. The results are shown in Figure \ref{fig4}. Red symbols represent bursts with $T_{\rm B}\geq 10^{33}\,\rm K$ within these two samples and it is obvious that all these red ones lie fairly close to the best-fitting line. As a comparison, grey dots are those with $T_{\rm B}<10^{33}\,\rm K$ and lie farther away from the line. This greatly increase the confidence of our classification by $T_{\rm B,cri}$.

\begin{figure*}
	\begin{center}
		\includegraphics[width=\textwidth]{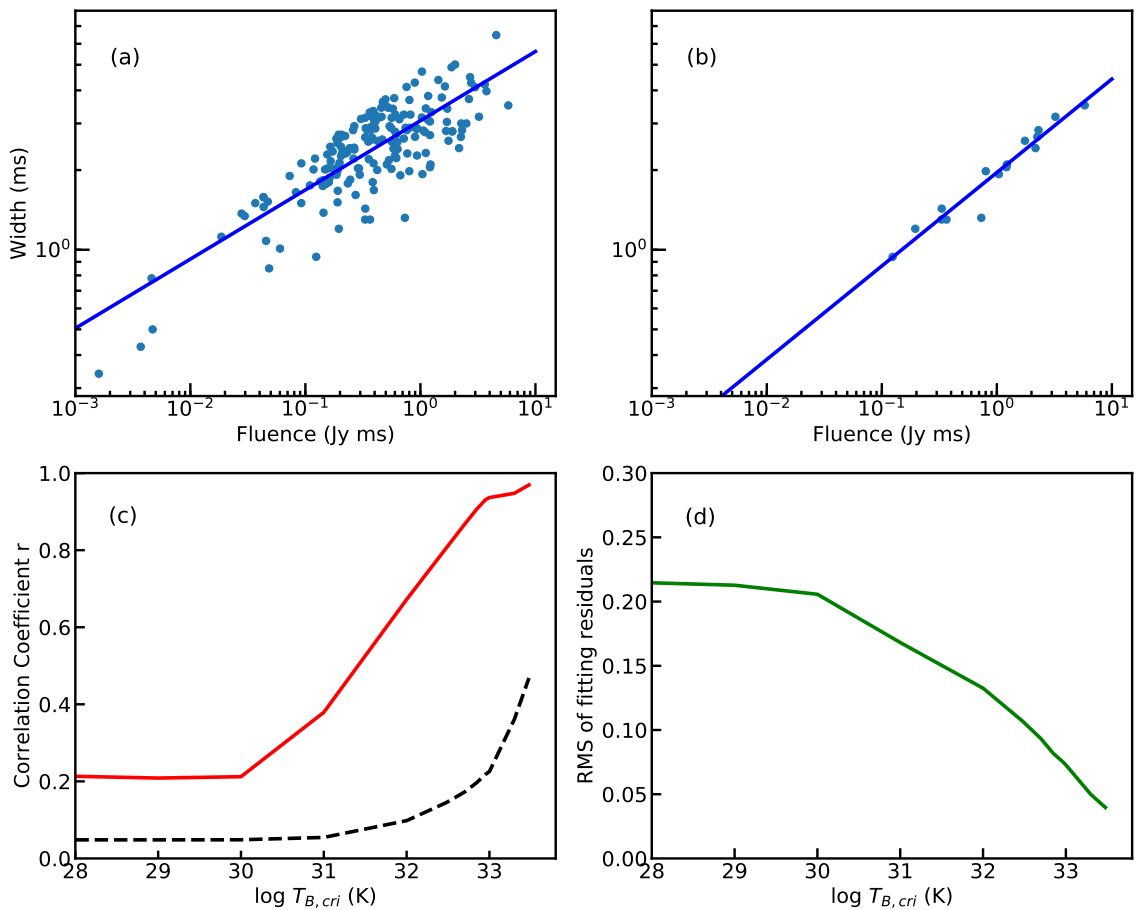}
		\caption{Panel(a): The case of $T_{\rm B,cri}=10^{32.5}\,\rm K$. Total 181 bursts with $T_{\rm B}\geq T_{\rm B,cri}$ indicate $T\propto\mathcal{F}_\nu^{0.261}$. Panel(b): The case of $T_{\rm B,cri}=10^{33.5}\,\rm K$. Total 16 bursts with $T_{\rm B}\geq T_{\rm B,cri}$ indicate $T\propto\mathcal{F}_\nu^{0.353}$. Panel(c):
The calculated correlation coefficient versus $T_{\rm B,cri}$ (red solid line). The black dashed line represents the critical $r$ for null hypothesis. Panel(d): The RMS of fitting residuals versus $T_{\rm B,cri}$.}
		\label{fig5}
	\end{center}
\end{figure*}

However, one issue remains that we choose $T_{\rm B,cri}\simeq10^{33}\,\rm K$ somewhat subjectively. Other values of $T_{\rm B,cri}$ should also be tested for completeness. We then adopt different values and check whether the $T-\mathcal{F_\nu}$ relation still holds. We have plotted another two cases of $T_{\rm B, cri}=10^{32.5},10^{33.5}\,\rm K$ in panel(a)(b) of Figure \ref{fig5} respectively. For relatively low $T_{\rm B,cri}$, the scatter is very large and the correlation is loose. Generally, the correlation coefficient $r$ gets larger for stronger correlation between two quantities. We have plotted its evolution with $T_{\rm B,cri}$ in panel(c). Red solid line represent the calculated $r$ and black dashed line represent the critical value of rejection regions. The fact that the red line is above the black one means $T-\mathcal{F_\nu}$ correlation is always expected. Clearly we see that $r$ increases with $T_{\rm B,cri}$, indicating that the relation becomes tighter for higher $T_{\rm B,cri}$. Moreover, we have also plotted the dependence of root mean square (RMS) of fitting residuals on $T_{\rm B,cri}$ in panel(d). The RMS decreases as $T_{\rm B,cri}$ increases. This makes sense since it is easier to pick out real ``classical'' bursts with higher criterion on $T_{\rm B}$. Around $10^{33}\,\rm K$, this relation is prominent enough and it is reasonable to take this value for FRB 20121102A. Numerical estimation gives the similar value later in Section \ref{sec4}. Note that $T_{\rm B,cri}$ could differ significantly for other FRB events, which will be discussed later.

\section{Discussion and Conclusions}
\label{sec4}
This paper has presented a data-oriented research and we have discussed a promising criterion for classifying FRBs, that is the brightness temperature. We have analyzed the FAST sample of FRB 20121102A and found that $T_{\rm B, cri}\simeq10^{33}\,\rm K$ is a reasonable dividing line. Bursts with $T_{\rm B}\geq T_{\rm B, cri}$ are ``classical'' FRBs. For this subtype we have found a tight relation $T\propto\mathcal{F_\nu}^{0.306}$. Further this relation has been verified using different samples.

According to Eq.(\ref{eq:T_B}), $T_{\rm B}$ is roughly in proportion to $\mathcal{F}_\nu/T^3$, then we can get a relation $T \sim \mathcal{F}_{\nu}^{1/3}$ directly for a fixed brightness temperature. However, this simple scaling could not be extended to a sample of bursts with a wide range of $T_{\rm B}$, as we can clearly see in the upper panel of Figure 3, in which no clear relation exists for the total 1652 FAST bursts. The positive power-law relation is hidden because the full sample is a mixture of two types of bursts, and finally emerges if we separate them by  $T_{\rm B,cri}$. The width-fluence relation we found is actually for bursts with $T_{\rm B}$ higher than some critical value, i.e., $T(T_{\rm B}\geq 10^{33} \,{\rm}K)\propto \mathcal{F}_{\nu}^{0.306}$.

\begin{figure}
	\begin{center}
		\includegraphics[width=0.42\textwidth]{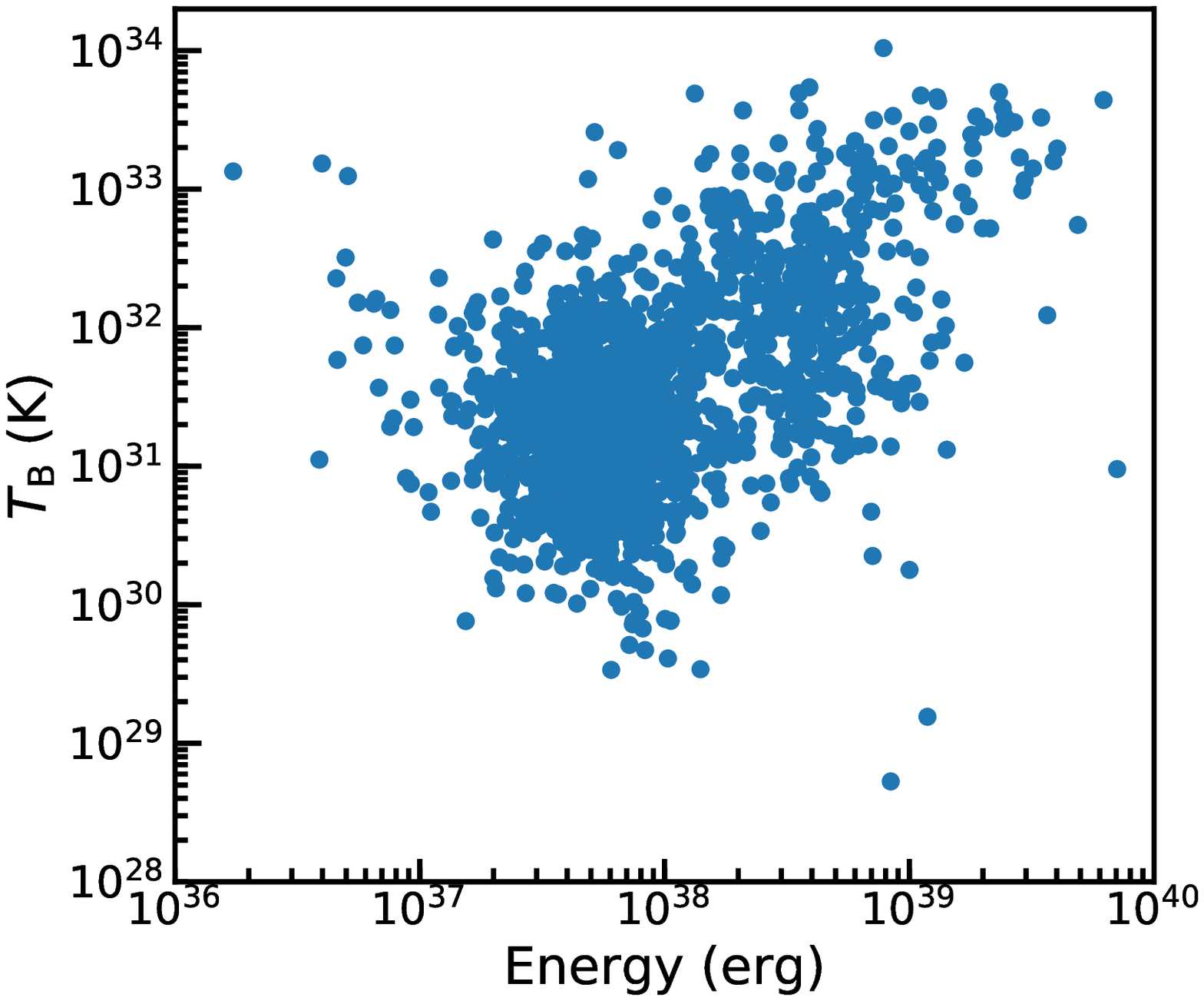}
		\caption{The brightness temperature versus burst energy for the full FAST sample.}
		\label{fig6}
	\end{center}
\end{figure}

A temporal energy distribution is clear in Figure 1 of \cite{LiD2021}, favoring the argument that two types of bursts exist. As we can infer from Eq.(\ref{eq:T_B}), $T_{\rm B}\propto E_{\rm FRB}/T^3$. For bursts with higher energy, their brightness temperature is generally higher, as we can see in the trend of Figure \ref{fig6}. The temporal energy distribution actually reflects the rate of two burst types. There are 1576 ``atypical'' bursts in the FAST sample after our classification. The majority of these bursts should have a different radiation mechanism from ``classical'' FRBs, for which the brightness temperature could hardly reach $T_{\rm B, cri}$. These bursts could be pulsar radio emission, giant pulses or some other kind of radio emission that have a much higher burst rate than ``classical'' FRBs. Actually if we plot 1576 bursts on the radio transient phase space, they fill the gap between
pulsar radio emission and FRBs. Note that the Galactic FRB 20200428A also lies in this range \citep{Bochenek2020,CHIME2020b,Kirsten2020}, thus could also be regarded as ``atypical'' burst. Alternatively, it is possible that these ``atypical'' bursts may not belong to a single subtype and consist of different types of radio transient, therefore the correlation between $T$ and $\mathcal{F_\nu}$ is very loose.

The existence of the dividing line should be related to FRB radiation mechanism directly. In principle, the emission frequency, instead of the central frequency of the receiver, should be used to obtain a more physical $T_{\rm B}$ value. The reason is that some bursts are narrow-banded, for instance, a sample of low-energy bursts from FRB 20121102A are generally detected in less than one-third of the observing bandwidth \citep{Gourdji2019}. Moreover, the pulse width in Eq.(\ref{eq:T_B}) should be intrinsic width. Because a physical dividing line, if exist, should be totally intrinsic to the source, and not affected by propagation effect and observation bias. This dividing line is naturally expected, since different types of coherent radio emission have been found to be separated by their brightness temperature in radio transient phase space \citep{Nimmo2021}.

The radiation mechanism of FRBs is largely unknown. Currently, two leading theories are coherent curvature emission by bunched particles in the magnetosphere of a neutron star, and synchrotron maser emission from magnetized shocks outside the magnetosphere \citep[for reviews, see][]{Zhang2020c,XiaoD2021}. Constraints can be made using the critical value of brightness temperature. We can rewrite Eq.(\ref{eq:T_B}) as
\bea
kT_{\rm B}=\frac{\nu F_\nu d_{\rm A}^2}{2\pi\nu^3T^2}\simeq\frac{L}{8\pi^2\nu^3T^2}.
\ena
Therefore, the FRB luminosity can be estimated as
\bea
L=1.09\times10^{40}\,{\rm erg\,s}^{-1}\nu_{\rm GHz}^3 T_{\rm ms}^2T_{{\rm B},33}.
\label{eq:LFRB}
\ena
For the coherent curvature emission mechanism, its luminosity can be expressed as \citep{Zhang2021}
\beq
L_{\rm curv}=N_{\rm b}N_{\rm e}^2\gamma^2P_{\rm curv},
\label{eq:Lcurv}
\enq
where $N_{\rm b}$ is the number of bunches, $N_{\rm e}$ is the number of electrons in one bunch and $\gamma$ is the Lorentz factor of radiating electrons. The radiation power of a single electron is
\beq
P_{\rm curv}=\frac{2}{3}\frac{\gamma^4e^2c}{\rho^2}\simeq4.61\times10^{-15}\,{\rm erg\,s^{-1}}\gamma_{2.5}^4\rho_8^{-2}.
\label{eq:Pcurv}
\enq
where $\rho$ is the curvature radius. The length of the bunch should be smaller than emission wavelength $\lambda$ in order to be coherent, and the transverse size of causally connection can be approximated as $\sim\gamma\lambda$ \citep{Kumar2017}. Therefore
\bea
N_{\rm e}\simeq\mathcal{M}n_{\rm GJ}\pi\gamma^2\lambda^3=5.89\times10^{17}\mathcal{M}\gamma_{2.5}^2\nu_{\rm GHz}^{-3},
\label{eq:Ne}
\ena
where $\mathcal{M}$ is the multiplicity factor and the Goldreich-Julian number density is \citep{Goldreich1969}
\bea
n_{\rm GJ}=\Omega B/(2\pi e c)=6.94\times10^7\,{\rm cm^{-3}}B_{{\rm s},15}P^{-1}R_8^{-3},
\label{eq:nGJ}
\ena
where we assume a magnetar engine with surface magnetic field $B_{\rm s}\sim10^{15}\,\rm Gauss$, rotation period $P\sim 1\,\rm s$ and emission radius $R\sim10^8\,\rm cm$. Substituting Eq.(\ref{eq:Pcurv})(\ref{eq:Ne}) into Eq.(\ref{eq:Lcurv}), we have
\bea
L_{\rm curv}\simeq1.60\times10^{31}{\,\rm erg\,s^{-1}}N_{{\rm b},5}\mathcal{M}^2\gamma_{2.5}^{10}\nu_{\rm GHz}^{-6}\rho_8^{-2}.
\label{eq:Lcurvvalue}
\ena
Basically, the Lorentz factor $\gamma$ can not exceed $10^3$ due to the drag force exerted by resonant scattering with the soft X-ray radiation field of the magnetar \citep{Beloborodov2013a}. The multiplicity $\mathcal{M}\sim10^2$ from resonant scattering is expected \citep{Beloborodov2013b}. Comparing Eq.(\ref{eq:LFRB})(\ref{eq:Lcurvvalue}), we find that in the most optimistic case, the brightness temperature of coherent curvature emission can merely reach $10^{33}$ K. For $T_{\rm B}>10^{33}\,\rm K$, this model requires unrealistic large value of $N_{\rm b}N_{\rm e}^2$, for which the formation and maintenance of the bunches can be problematic \citep{Saggion1975,Cheng1977, Kaganovich2010}. A more stringent constraint on $T_{\rm B}$ is given based on a similar method \citep{Lyutikov2021b}. However, this mechanism can still responsible for ``atypical'' bursts.

For the maser mechanism from magnetized shocks, a very small fraction of shock energy is released in the form of radio precursor \citep{Hoshino1991}. Particle-in-Cell simulation suggests that this fraction is $\sim7\times10^{-4}/\sigma_{\rm u}^2$, where $\sigma_{\rm u}\geq1$ is the upstream magnetization \citep{Plotnikov2019}. Moreover, only a fraction of bolometric maser power can turns into FRB emission due to strong induced Compton scattering \citep{Metzger2019}. These two effects give a combined fraction $f\sim10^{-6}-10^{-5}$ \citep{XiaoD2020}, so that
\bea
L_{\rm maser}\sim f E_{\rm flare}/T\sim 10^{37}\,{\rm erg\,s^{-1}}f_{-6}E_{{\rm flare},40}T_{\rm ms}.
\label{eq:Lmaser}
\ena
where the flare energy $E_{\rm flare}$ is assumed to be comparable to magnetar X-ray bursts energy $E_X$. Observations of hundreds of X-ray bursts from SGR 1806-20 and SGR 1900+14 gives that $E_X\sim10^{38}-10^{41}\,\rm erg$ \citep{Gogus1999,Gogus2000}. Comparing Eq.(\ref{eq:LFRB})(\ref{eq:Lmaser}), we find maser emission from shocks by normal flares can not produce high brightness temperature FRBs. Magnetar giant flares with $E_{\rm flare}>10^{43}\,\rm erg$ is need to reach $T_{\rm B}>10^{33}\,\rm K$. However, giant flares can not happen as frequently as FRB 20121102A bursts. Also, the synchrotron maser mechanism has difficulty in explaining polarization swing \citep{Luo2020} and nano-second variability in the light curve \citep{Lu2021}.

To conclude, both these two mechanisms have their own limitations. Conservatively speaking, both of them seem unlikely to produce FRBs with $T_{\rm B}>10^{33}\,\rm K$. So it remains unclear what mechanism makes the ``classical'' FRBs. There have been tens of possible mechanisms for pulsar radio emission, and it has been proposed that some of them (e.g., free electron laser mechanism) can be applied to FRBs and reach high brightness temperature \citep{Lyutikov2021a}. Interestingly, \citet{Zhang2021} recently suggested that the coherent inverse Compton scattering could be a promising mechanism, in which the bunch formation problem can be largely relieved due to enhanced emission power of a single electron. Plenty of works need to be done to figure out FRB radiation mechanism, and the $T-\mathcal{F_\nu}$ relation could provide some hints for future studies. Lastly, while we expect a critical brightness temperature for every FRB event, the difference in source property and environment could result in different $T_{\rm B, cri}$ value. A further analysis on other FRBs using similar method is now in progress. 

\begin{acknowledgements}
We would like to thank an anonymous referee for helpful comments. This work is supported by the National Key Research and Development Program of China (Grant No. 2017YFA0402600), the National SKA Program of China (grant No. 2020SKA0120300), and the National Natural Science Foundation of China (Grant No. 11833003, 11903018). DX is also supported by the Natural Science Foundation for the Youth of Jiangsu Province (Grant NO. BK20180324).
\end{acknowledgements}

\bibliographystyle{aa} 
\bibliography{FRBlatest} 

\end{document}